\documentclass{article}
\usepackage{spconf,amsmath,graphicx}
\usepackage{xcolor}
\usepackage{booktabs}
\usepackage{lipsum}
\usepackage{soul}
\usepackage{caption}
\usepackage{subcaption}
\usepackage{siunitx}
\usepackage{url, cite}
\usepackage{enumitem}
\usepackage[breaklinks]{hyperref}
\usepackage{booktabs}
\usepackage{float}
\usepackage{xurl}
\setlist[itemize]{noitemsep, topsep=1pt}

\title{Rate-Quality or Energy-Quality Pareto Fronts for\\ Adaptive Video Streaming?}

\name{Angeliki Katsenou, Xinyi Wang, Daniel Schien, and David Bull\thanks{This work has been supported by the UKRI Creative Bristol+Bath Cluster, the UKRI MyWorld Strength in Places Programme (SIPF00006/1), and the Bristol Digital Futures Institute.}}
\address{Visual Information Lab, School of Computer Science, University of Bristol, Bristol BS1 8UB, UK }

\begin{document}
%\ninept
%
\maketitle
\begin{abstract}
Adaptive video streaming is a key enabler for optimising the delivery of offline encoded video content. The research focus to date has been on optimisation, based solely on rate-quality curves. This paper adds an additional dimension, the energy expenditure, and explores construction of bitrate ladders based on decoding energy-quality curves rather than the conventional rate-quality curves. Pareto fronts are extracted from the rate-quality and energy-quality spaces to select optimal points. Bitrate ladders are constructed from these points using conventional rate-based rules together with a novel quality-based approach. Evaluation on a subset of YouTube-UGC videos encoded with x.265 shows that the energy-quality ladders reduce energy requirements by 28-31\%  on average at the cost of slightly higher bitrates. The results indicate that optimising based on energy-quality curves rather than rate-quality curves and using quality levels to create the rungs could potentially improve energy efficiency for a comparable quality of experience.
\end{abstract}
%

%%%%%%%%%%%%%%%%%%%%%%%%%%%%%%%%%%%%%%%%%%%%%%%%%%%%%%%%%%%%%%%%%

\section{Introduction}
\label{sec:intro}
The immense growth in the consumption of video data has been associated with the increased demand for ``better'' and more ``realistic'' content, alongside the exploitation of  higher spatial and temporal resolutions, higher dynamic range and more immersive formats~\cite{sandvine}. The present scale of global video delivery via streaming demands improved compression technologies capable of achieving  increased compression ratios while improving the reconstruction quality compared to previous technology generations. However, this also results in higher computational complexities that are directly linked to higher energy consumption across the whole streaming pipeline. 
% It is worth noting that the global greenhouse emissions attributed to video services was estimated at about 1.3\% in 2015~\cite{CarbonImpactVS} and the expectation is that this has increased.

As video encoding/decoding are energy-intensive processes, an important step towards the design of interventions to reduce their  carbon footprint is first to understand the codec energy consumption.
Previous research has experimented with the energy profiling of encoding/decoding for different standards, such as H.265, VVC, and AV1~\cite{KatsenouPCS2024, KatsenouPCS2022, Chachou_MMSP2023, HerglotzCSVT2019, PakdamanICIP2020, KraenzlerCSVT2021}, either using software estimators of power such as Intel's RAPL~\cite{RAPL2018} or hardware-based power meters, such as Tektronix PA1000~\cite{KatsenouPCS2024}.
Moreover, research has also focused on assessing the energy consumed at decoding on different end user devices, e.g., phones, laptops, displays~\cite{KraenzlerCSVT2022, RaakeQoMEX2022}.

Further to energy profiling, potential solutions to increase energy efficiency have been proposed. For example, Herglotz et al.~\cite{Herglotz_QoMEX2023} studied the effect of display choice and configuration settings on energy consumption. Amirpour et al.~\cite{AmirpourICME2023}, explored the energy savings across different x.265 encoding presets and proved that the choice of preset can significantly impact both the quality and energy consumption of video encoding. 

Technical solutions can increase the efficiency of services by reducing the \textit{relative} required input of energy for an desired level of quality. Additionally, the concept of ``sufficiency'' has been proposed to refer to strategies that directly aim for \textit{absolute} impact reductions from lowering production and consumption~\cite{Santarius2022}. Sufficiency can be translated across different layers of a digital service: hardware, software, user, and economic. In our case, we consider ``software sufficiency'' which here relates to codec settings and parameterisation, and ``user sufficiency'' that represents a satisfactory quality of viewing experience~\cite{KatsenouMMTC2023}. Within this context, Bingol et al.~\cite{Bingol_ICC2023} explored whether limiting the maximum Quality of Experience (QoE) (translated in Structure SIMmilarity index (SSIM)~\cite{Bovik_SSIM} values) to an acceptable level at a fixed set of parameters per display device would be a possible solution to reduce energy consumption while still satisfying consumer quality expectations. Further to that, efforts started accounting for the energy consumption of dynamic adaptive streaming while maintaining a high quality of experience~\cite{Menon_VCIP2023, Menon_MHV2024}. Menon et al.~\cite{Menon_VCIP2023} followed a cross-codec approach and the basic idea lied in the elimination of representations within similar perceptual quality range (across codec curves). Selecting representations of similar quality resulted in energy savings. Recently, Menon et al.~\cite{Menon_MHV2024} offered another solution through the incorporation of spatial resolution prediction models that reduce the number of required encodes, thus the energy consumption~\cite{Menon_MHV2024}.

In this work, we attempt to apply the concept of sufficiency of the user experience, bundled with the energy consumption within an adaptive streaming scenario. We ask the question of whether Rate-Quality (RQ) optimisation remains the best solution to build bitrate ladders for adaptive video streaming when taking into consideration the energy consumption. Therefore, besides the traditional RQ method as in our previous work~\cite{KatsenouOJSP2021, KatsenouPCS2021}, we explore the tradeoffs when constructing bitrate ladders based on the Energy-Quality (EQ) curves. To the best of our knowledge, this has not been explored before. To represent subjective quality, we consider the Video Multimethod Assessment Fusion (VMAF)~\cite{VMAFblog} metric, as it aligns well with human visual perception and is widely used by in the research community and the industry.

The remainder of this paper is organised as follows. Understanding the parameter space is important, therefore the dataset and measurements utilised are presented in Section~\ref{sec: ExperimentalSetup}. Section~\ref{sec:RQELadders} presents the methodology to create the RQ and EQ Pareto Fronts (PFs) and the construction of the ladders. Next, Section~\ref{sec:EvaluateLadders} discusses the evaluation results. Finally, conclusions are drawn in Section~\ref{sec:Conclusion}.

%%%%%%%%%%%%%%%%%%%%%%%%%%%%%%%%%%%%%%%%%%%%%%%%%%%%%%%%%%%%%%%%%
\section{Test sequences, Video Codec, and Energy Measurements}
\label{sec: ExperimentalSetup}
Prior to explaining  the details of our work, we briefly present the experimental process used to compute the energy expended and the quality delivered after encoding a set of video sequences.
We selected sequences from the YouTube-UGC~\cite{AdsumilliMMSP2019} dataset, as it comprises different genres and is representative of streamed user generated content. 
A variety of content is crucial when testing algorithms/processes related to compression, as their performance is content-dependent. Videos with complex motion patterns or/and dynamic textures are harder to compress~\cite{KatsenouPCS2016b}. Such videos are expected to require higher energy for both encoding and decoding. 
All 2160p native sequences from the Animation, Gaming, Sports, HDR, and Vlogs genres were selected (82 in total). Most videos have a YUV 4:2:0 color sampling except for the HDR examples that have a YUV 4:2:2 format. The video duration is \SI{20}{sec}, however the frame rates vary, ranging from 15 to \SI{60}{fps}\footnote{A full list of the sequences used along with the power measurements is available on the project page~\cite{ProjectPage_Xinyi}.}.

For the encoding of the test sequences, we used the ffmpeg N-110021-g85b185b504 version~\cite{ffmpeg} implementation of H.265/HEVC~\cite{r:HEVC, j:Ohm, b:Wien}. From the variety of presets, we selected the default, medium, in the Constant Rate Factor (CRF) mode, which allows consistent quality across frames. We used five CRF values, $\{10, 20, \ldots, 50\}$ to capture the whole range of quality-rate-energy tradeoffs. Our power measurements rely on the integrated power meter in Intel CPUs, the RAPL~\cite{RAPL2018}. RAPL reports the energy consumption on different levels or power domains: entire CPU socket, all CPU cores, integrated graphics, and dynamic random-access memory (DRAM). RAPL has been frequently used in similar research activities~\cite{KatsenouPCS2024, KatsenouPCS2022, Chachou_MMSP2023}. The workstation utilised for the compression experiments has an Intel(R) Core(TM) i9-7900X CPU @\SI{3.30}{GHz} and \SI{64}{GB} RAM.

%%%%%%%%%%%%%%%%%%%%%%%%%%%%%%%%%%%%%%%%%%%%%%%%%%%%%%%%%%%%%%%%%
\section{Exploring Adaptive Streaming based on Quality-Energy curves}
\label{sec:RQELadders}

\begin{figure} [!t]   
      \centering
      \includegraphics[width=8.6cm]{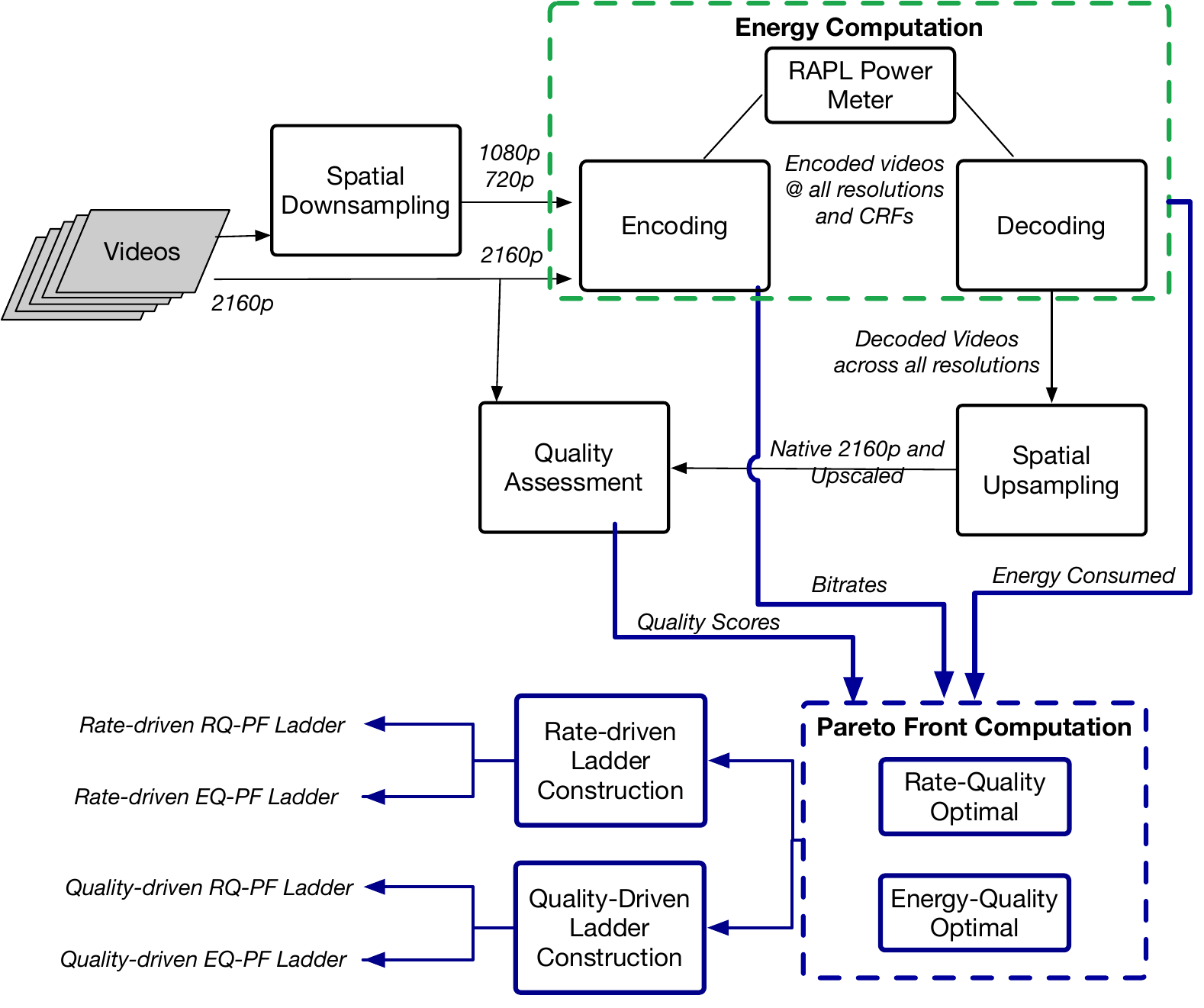}
    \caption{Overview of the proposed methodology. Black denotes the typical process of pre-processing for adaptive video streaming, green the energy computation as in~\cite{KatsenouPCS2024}, and blue the proposed.}
    \label{fig: Overview}
    \vspace{-1.5em}
\end{figure}

In order to create bitrate ladders, we need to find the PFs of the RQ and EQ space. To this end, as depicted in Fig.~\ref{fig: Overview}, we follow a similar pipeline to that reported in our previous work~\cite{KatsenouPCS2021}. We first downscale the native 2160p sequences to 1080p and 720p using a three-tap Lanczos~\cite{Duchon} filter. Next, we encode at different compression levels and then decode. These are the two processes that are probed separately to be measured for the power consumption and energy calculation with the RAPL power meter~\footnote{In this study, we do not include the energy consumption from the display. A detailed description of the measurement methodology can be found in~\cite{KatsenouPCS2024}}. After decoding, the 1080p and 720p sequences are upscaled to 2160p for the computation of the VMAF quality metric. These rate-energy-quality values are used to explore the parameter space and compute the PFs, as follows:
\begin{itemize}[leftmargin=*] 
  \setlength{\itemsep}{1pt}
  \setlength{\parskip}{1pt}
  \setlength{\parsep}{0pt}
    \item[-] \textit{RQ optimal}: this is the conventional approach that uses the RQ curves across all spatial resolutions to extract the RQ-PF.
    \item[-] \textit{EQ optimal}: considering the EQ curves (instead of RQ) across all spatial resolutions, the EQ-PF is extracted.
\end{itemize}%\vspace*{-\baselineskip}
From a recent analysis from the carbon Trust~\cite{CarbonTrust2021}, user devices are driving the environmental impact of video streaming due to the high number of viewers. Therefore, we base our exploration on the decoding energy.

\begin{figure} [!t]   
    \begin{minipage}[b]{.48\linewidth}
      \centering
      \centerline{\includegraphics[width=4.5cm]{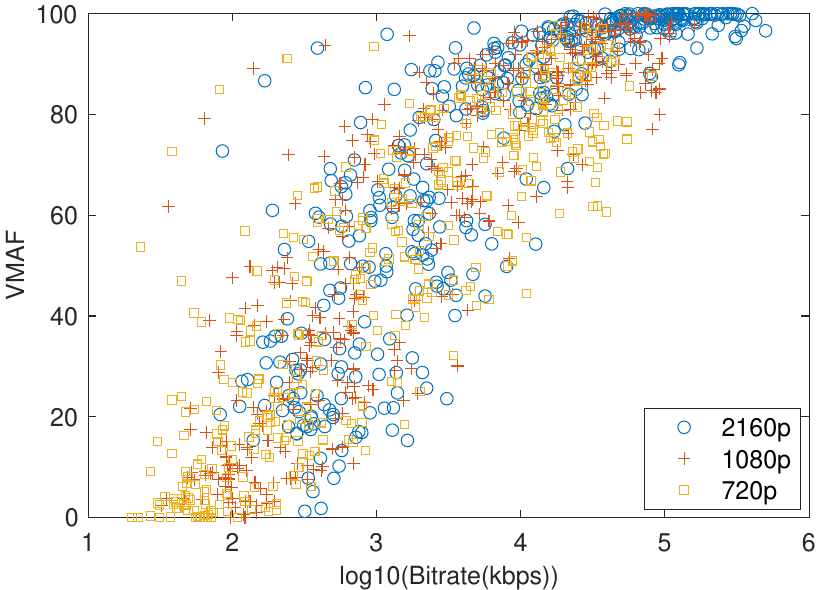}}
    %  \vspace{1.5cm}
      \centerline{(a) RQ parameter space.}\medskip
    \end{minipage}
    \hfill
    \begin{minipage}[b]{0.48\linewidth}
      \centering
      \centerline{\includegraphics[width=4.5cm]{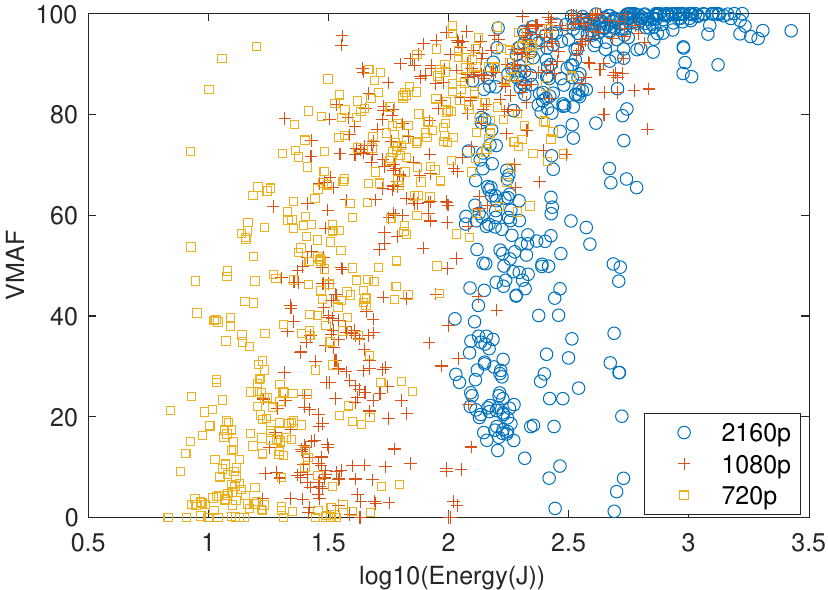}}
    %  \vspace{1.5cm}
      \centerline{(b) EQ parameter space.}\medskip
    \end{minipage}
    \caption{The two figures illustrate the quality-rate-energy points for encodes with x.265 across three spatial resolutions: 2160p, 1080p, and 720p. The energy here refers to the energy consumed during the decoding process.}
    \vspace{-1em}
    \label{fig: RQE parameter space}
\end{figure}

\subsection{Rate-Quality-Energy Parameter Space}
\label{ssec:RQE}
As described above, we compress video sequences at different compression levels and record the bit rate, quality, and energy. 
Figure~\ref{fig: RQE parameter space} illustrates the RQ and EQ parameter space across the three spatial resolutions for all tested sequences. Instead of bitrate and energy we used the logarithm base 10, to reduce the density of the points. In Fig.~\ref{fig: RQE parameter space}(a), as expected we notice an overlap in bitrate and quality across spatial resolutions. This overlap is increased compared to previous reports~\cite{KatsenouPCS2022}, attributed to the use of UGC content that is already pre-compressed, rather than professional content. In Fig.~\ref{fig: RQE parameter space}(b), the VMAF values are plotted against the decoding energy across all three resolutions. The horizontal shift in the decoding energy consumption range across the three resolutions can be easily observed. Taking into account that, in the RQ domain,  the bitrate ranges overlap significantly across the three resolutions, this indicates that a new  approach to construction of the bitrate ladder could provide benefits in terms of energy consumption. Another important observation is that, for decoding at high compression rates where the VMAF values are below 60, the energy consumption is of comparable range within each resolution group.

\subsection{Rate-Quality and Energy-Quality Pareto Fronts}
\label{ssec:RQEPFs}

\begin{figure} 
     \begin{minipage}[b]{.47\linewidth}
      \centering
      \ \subcaptionbox{Histogram of \% spatial resolutions for the RQ-PF.}{\includegraphics[width=4.5cm]{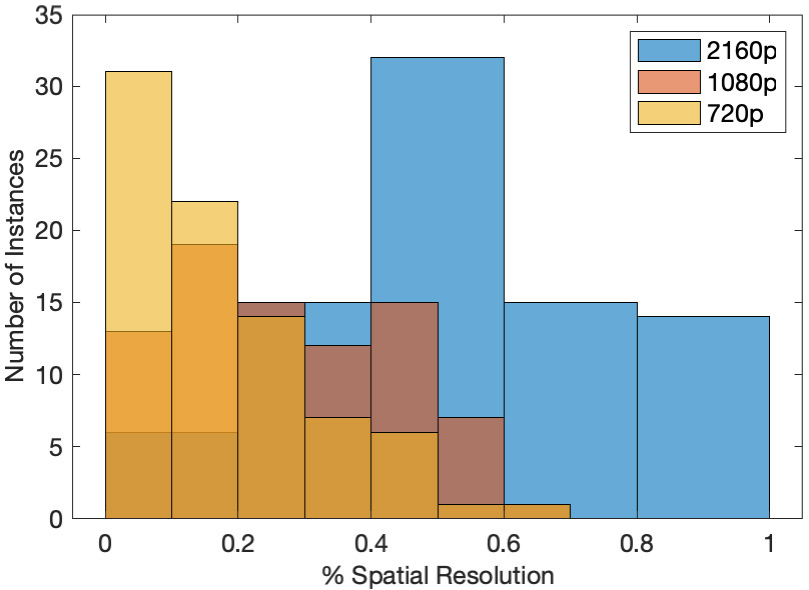}}
    \end{minipage}
    \hfill
    \begin{minipage}[b]{0.47\linewidth}
      \centering
      \subcaptionbox{Histogram of \% spatial resolutions for the EQ-PF.}{\includegraphics[width=4.5cm]{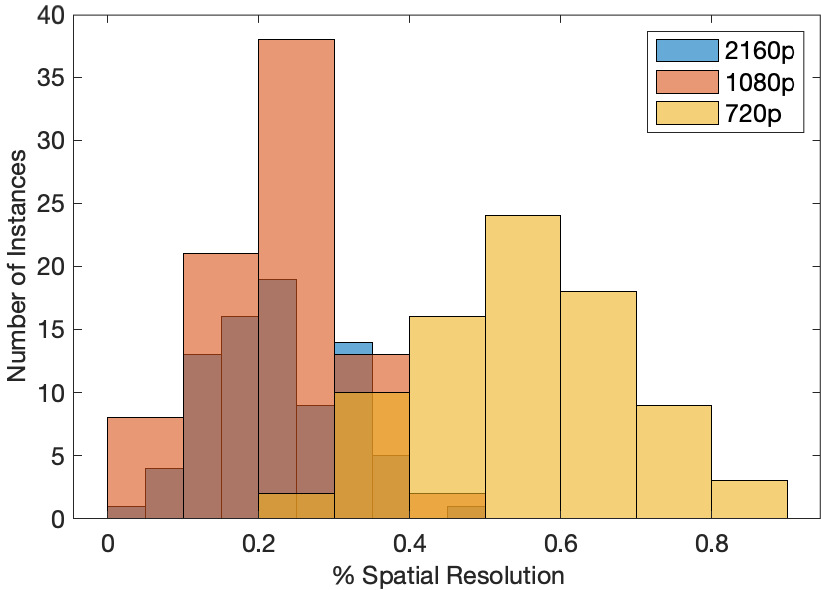}}
    \end{minipage}
    \begin{minipage}[b]{.47\linewidth}
      \centering
      \subcaptionbox{PFs in the RQ domain.}{\includegraphics[width=4.5cm]{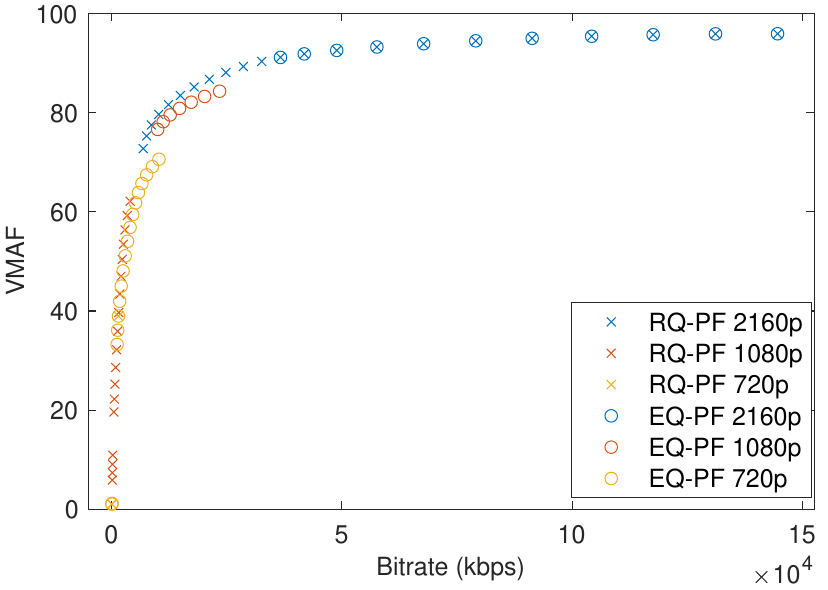}}
    \end{minipage}
    \hfill
    \begin{minipage}[b]{0.47\linewidth}
      \centering
      \subcaptionbox{PFs in the EQ domain.}{\includegraphics[width=4.5cm]{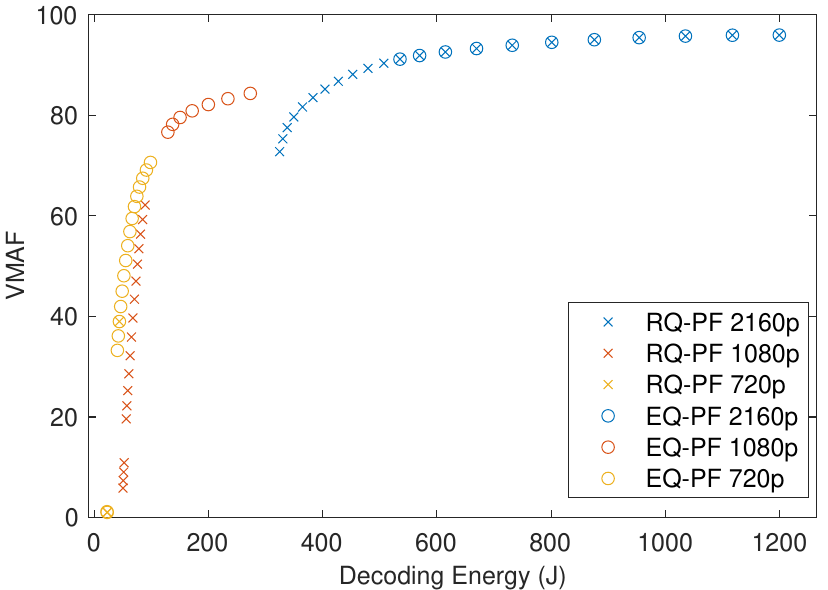}}
    \end{minipage}
   \caption{This figure summarises the findings in the energy-driven PF computation. The two histograms show the share of different spatial resolutions on the PFs. The bottom two plots illustrate the RQ-PF and EQ-PF for the test sequence ``Gaming\_2160P-67b0'' in the (c) RQ domain and (d) EQ domain. }
   \vspace{-1em}
    \label{fig: RQE-PFs}
\end{figure}

To construct the PFs for each sequence, we first apply Akima interpolation at all three dimensions, rate-quality-energy. Then, we select the PF points of the RQ and EQ curves. Those points are those that maximise quality for the minimum rate and energy, respectively. It is worth noting that the RQ-PF and the EQ-PF do not comprise the same set of points. A different combination of representations across spatial resolutions and CRF points comprise the RQ-PF and EQ-PF PFs. This composition is content-dependent; however on average for the considered dataset, 55\% of the RQ-PF are 2160p representations, while only 22\% for the EQ-PF. Typically, as illustrated in the histograms of Fig.~\ref{fig: RQE-PFs}(a)-(b), for the EQ-PF, a higher number of lower spatial resolution points are selected compared to the RQ-PF.

In Fig.~\ref{fig: RQE-PFs}(c)-(d), examples of EQ-PFs and RQ-PFs are illustrated for the ``Gaming\_2160P-67b0'' video in the RQ and EQ space. A first observation is that the projection of EQ-PF into the RQ domain results in a non-monotonic PF and vice versa. A second observation is that the RQ-PF consists in its majority of 2160p and 1080p representations, while the EQ-PF spans across all three resolutions. Also in this example it appears that, for VMAF lower than 50 (high compression) the energy consumption is very similar for the two PFs. However, for VMAF values within the range of 60 to 85, although the range of bitrates between the two PFs is overlapping, the range of energy consumption is significantly higher for the conventional RQ-driven method. In this case, for the EQ-driven solution, representations of lower spatial resolution were selected. Last, for VMAF values higher than 90, both methods perform similarly. A similar pattern has been observed for many other sequences.

\begin{figure}[!h]
    \begin{minipage}[b]{.47\linewidth}
      \centering
      \subcaptionbox{Rate-driven ladder construction at the RQ domain.}{\includegraphics[width=4.5cm]{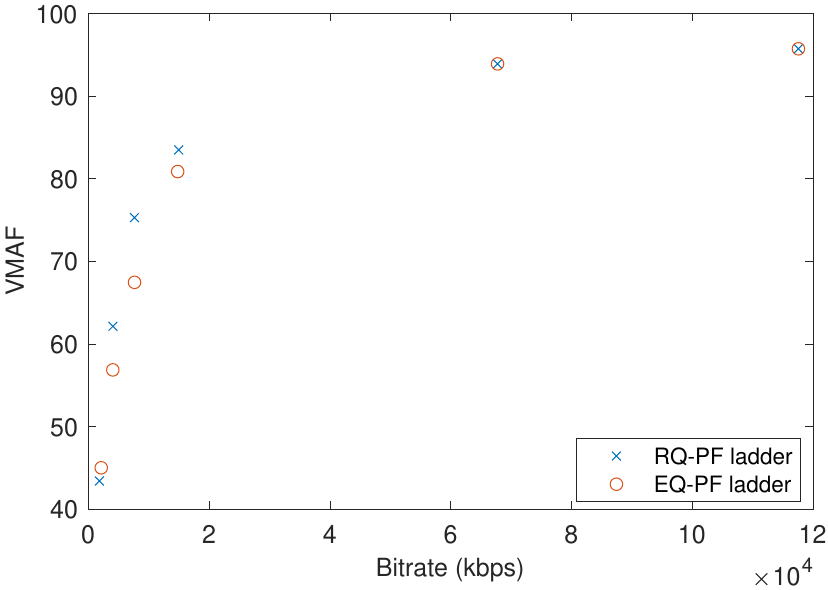}}
    \end{minipage}
    \hfill
    \begin{minipage}[b]{0.47\linewidth}
      \centering
      \subcaptionbox{Rate-driven ladder construction at the EQ domain.}{\includegraphics[width=4.5cm]{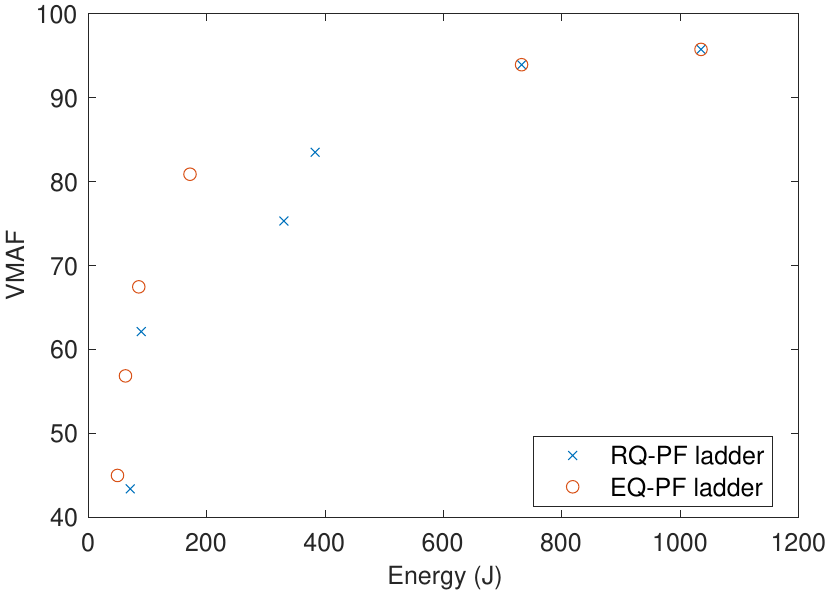}}
    \end{minipage}
    \begin{minipage}[b]{.47\linewidth}
      \centering
      \subcaptionbox{Quality-driven ladder construction at the RQ domain.}{\includegraphics[width=4.5cm]{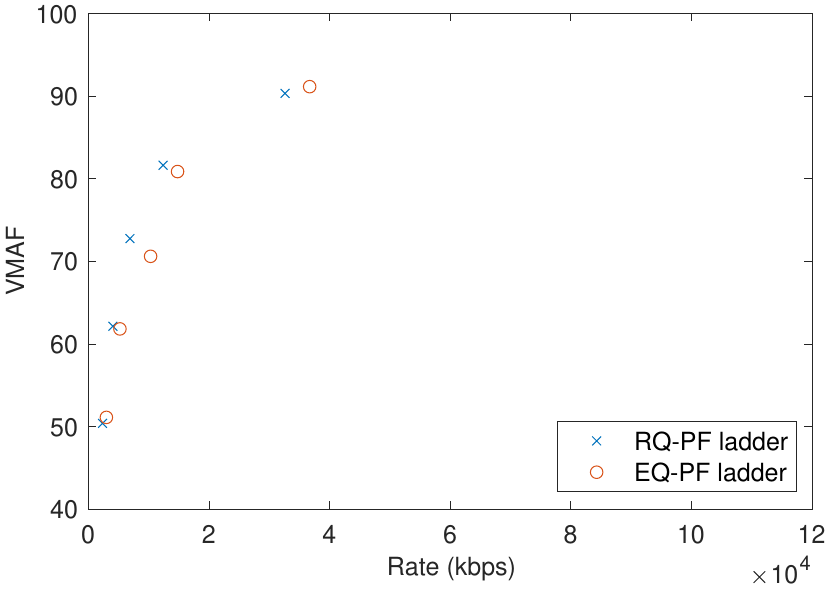}}
    \end{minipage}
    \hfill
    \begin{minipage}[b]{0.47\linewidth}
      \centering
      \subcaptionbox{Quality-driven ladder construction at the EQ domain.}{\includegraphics[width=4.5cm]{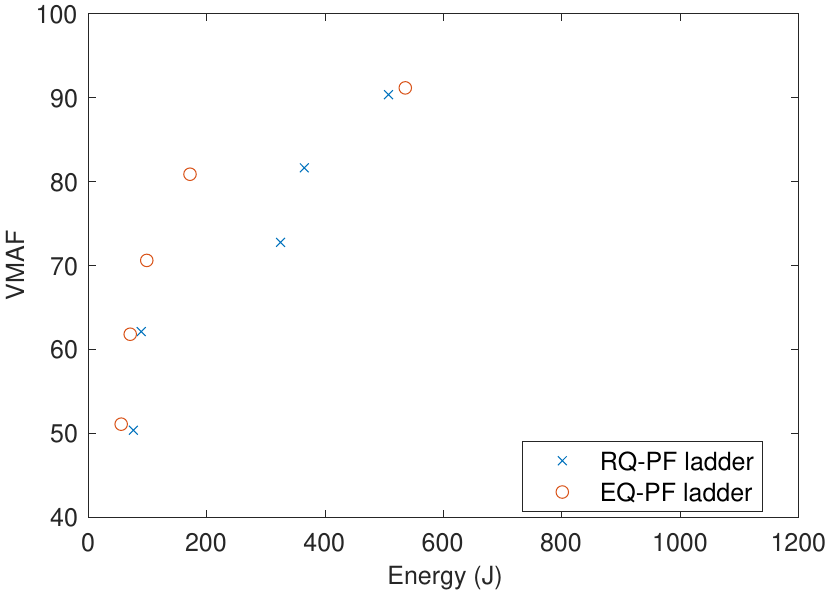}}
    \end{minipage}
   \caption{The two top plots illustrate the rate-driven RQ-PF and EQ-PF ladders for the test sequence ``Gaming\_2160P-67b0'' in the RQ and EQ domain, while the two bottom the quality-driven RQ-PF and EQ-PF ladders in the RQ and EQ domain.}
   \vspace{-1em}
    \label{fig: RQE-Ladders}
\end{figure}

\subsection{Building the RQ and EQ Ladders}
\label{ssec:RQEBuildLadders}

The final step in the proposed methodology is to build the bit rate ladders using the PFs produced by the  two methods and quantitatively compare them. Typically, in adaptive video streaming, RQ curves across different resolutions are used to extract the RQ-PF and then sample it at different bitrates $R_{L}$ to create the streaming ladder. In this work, we experiment with the following approaches:
\begin{itemize}[leftmargin=*] 
  \setlength{\itemsep}{1pt}
  \setlength{\parskip}{1pt}
  \setlength{\parsep}{0pt}
    \item[-] \textit{Rate-driven}: based on the conventional approach, using the two PFs, we sample based on the closest bitrate to the rung. A $10\%R_{L,i}$ range is considered to define the rung search area. In many cases, especially at lower bitrates, more than one representations from the PFs are within that range. In this case, the representation on the PF with the lowest bitrate within that range is selected. Another rule we applied, is that each new bitrate rung $R_{L,i}$ is twice that of the previous one, i.e. $R_{L,i}=2 R_{L,i-1}$, where $i\in{1, 2, \ldots, N}$ with $N$ the number of rungs. For the results presented in this work, we considered the [500kbps,128Mbps] bitrate range. The ladders constructed with this method are aligned in terms of bitrate but can vary a lot in terms of quality.
    \item[-] \textit{Quality-driven}: inspired by the quality ``sufficiency'', first we create quality rungs around certain quality levels. A $\Delta \textrm{VMAF} = 5$ is defining the range around the quality levels $Q_{L}$. We selected this value for $\Delta \textrm{VMAF}$ based on recent mapping of just noticeable distortion to VMAF scale~\cite{LeCallet_VMAFJND2022}. Finally, for the results presented, we considered the [50, 100] VMAF range in increments of $10$, i.e. $Q_{L,i}=Q_{L,i-1}+10$, where $i\in{1, 2, \ldots, N}$ with $N$ the number of rungs. The selected VMAF range is associated with medium to high quality. The ladders constructed with this method are aligned in terms of quality but can significantly vary in terms of bit rate.
\end{itemize}%\vspace*{-\baselineskip}
 
 Examples of the ladders created based on these methods on the ``Gaming\_2160P-67b0'' set are illustrated in Fig.~\ref{fig: RQE-Ladders}. As easily observed, the ladders produced by the rate-driven and the quality-driven methods are not identical for both the RQ-PF and EQ-PF ladders. The quality-driven method results in ladders that are limited within a narrower bitrate and energy range as this method allows only one representation per quality rung, eliminating representations of similar quality level. Furthermore, it is important to note the non-monotonicity of the RQ-PF ladder in the EQ domain.

%%%%%%%%%%%%%%%%%%%%%%%%%%%%%%%%%%%%%%%%%%%%%%%%%%%%%%%%%%%%%%%%%
\section{Results and Discussion}
\label{sec:EvaluateLadders}
All in all, we created the four ladders below, as shown in Fig.~\ref{fig: Overview}:
\begin{itemize}[leftmargin=20pt] 
  \setlength{\itemsep}{1pt}
  \setlength{\parskip}{1pt}
  \setlength{\parsep}{6pt}
    \item[i.] \textit{Rate-driven RQ-PF} ladder;
    \item[ii.] \textit{Rate-driven EQ-PF} ladder;
    \item[iii.] \textit{Quality-driven RQ-PF} ladder;
    \item[iv.] \textit{Quality-driven EQ-PF} ladder.
\end{itemize}

A typical way of comparing curves in the video compression domain is to use the Bj{\o}ntegaard delta metrics~\cite{r:Bjontegaard}. However, as explained in~\cite{barman2024bjontegaard}, this is not feasible with non-monotonic curves due to interpolation errors. As mentioned earlier, the produced ladders in many cases are non-monotonic, especially when projected onto a different domain. Specifically, RQ optimal ladders are often non-monotonic when projected to the EQ domain, and vice versa. Therefore, the Bj{\o}ntegaard delta metrics are not a suitable choice. Instead, we employ the mean relative difference to provide the quantitative comparison of the computed ladders:
\begin{equation}
    \delta_R = \dfrac{1}{N} \displaystyle \sum_{n=1}^N \dfrac{R_n^{Ref}-R_n^{Prop}}{R_n^{Ref}},
\end{equation}
where $N$ is the number of rungs, $R_n^{Ref}$ in the reference ladder (Rate-driven RQ-PF) rate point $n$ and $R_n^{Prop}$ is the rate point $n$ of the other explored solutions. Similarly, the relative difference of quality $\delta_{Q}$ and energy $\delta_{E}$ of the explored solutions is computed. The relative difference compensates for the difference in the order of magnitude of the measurements. The resulting relative difference values are reported in Table~\ref{tab: evalMetrics}. At both categories, Rate-driven and Quality-Driven, the RQ-PF is considered as the reference for the calculation of the metric. From this table it is evident, that the better RQ performance comes at the cost of higher energy.

\begin{table} [!h]
\caption{Mean value and standard deviation of the mean relative difference over rate, quality, and energy of the EQ-PF ladders against the RQ-PFs.}
\centering 
\resizebox{1.0\linewidth}{!}{
\begin{tabular}{l|c|c|c}
\toprule
Ladder &  $\overline{\delta}_{\textrm{rate}} \pm \sigma_{\delta_R}$ & $\overline{\delta}_{Q} \pm \sigma_{\delta_Q}$ & $\overline{\delta}_{E} \pm \sigma_{\delta_E}$\\
\midrule
Rate-driven EQ-PF&0.60\%$\pm$2.70\% & 4.35\%$\pm$3.55\% & 31.43\%$\pm$14.35\%\\
\midrule
Quality-driven EQ-PF& 34.46\%$\pm$34.33\%& 0.12\%$\pm$0.96\%& 28.23\%$\pm$19.08\%\\
\bottomrule
\end{tabular} }
\label{tab: evalMetrics}
 % \vspace{-.5em}
\end{table}

\begin{figure} [!t]   
    \begin{minipage}[b]{.46\linewidth}
      \centering
      \subcaptionbox{RQ domain.}{\includegraphics[width=4.4cm]{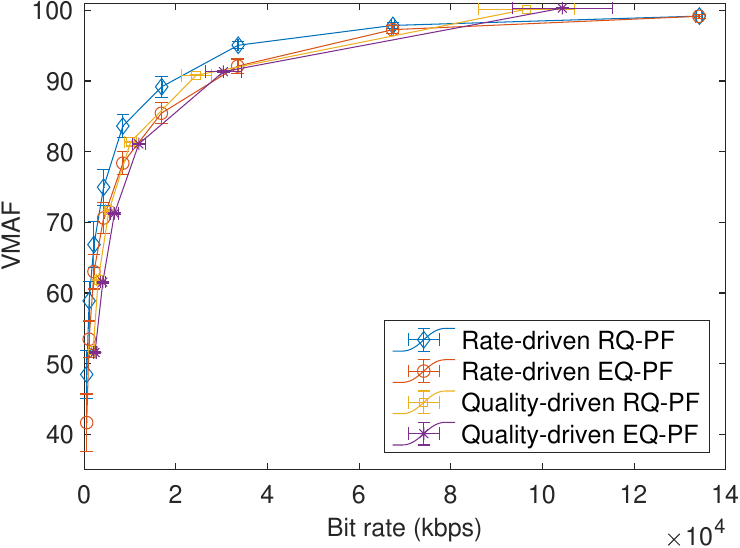}}
    \end{minipage}
    \hfil
    \begin{minipage}[b]{.48\linewidth}
      \centering
      \subcaptionbox{EQ domain.}{\includegraphics[width=4.5cm]{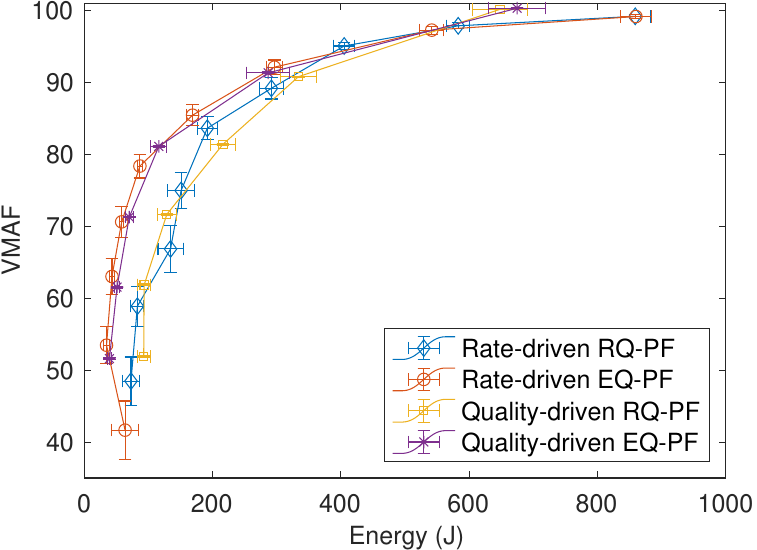}}
    \end{minipage}   
    \begin{minipage}[b]{\linewidth}
      \centering
      \subcaptionbox{RE domain.}{\includegraphics[width=4.6cm]{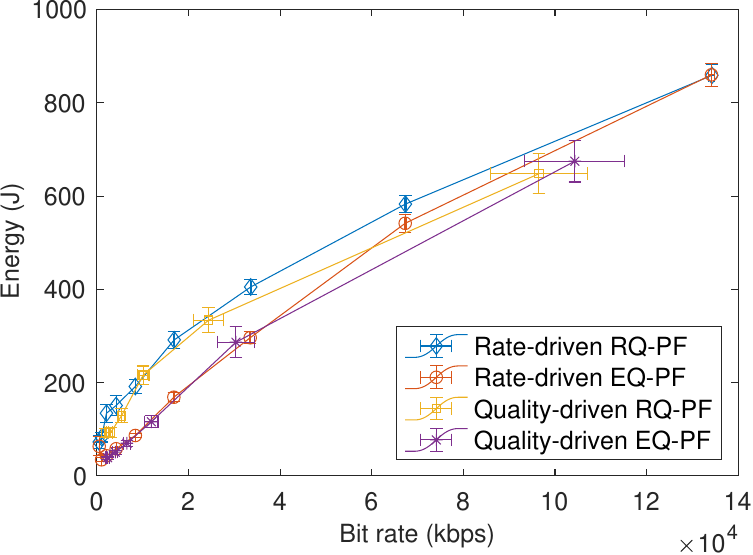}}
      \vspace{-.5em}
    \end{minipage}
    \caption{Mean Ladders with standard error over all videos.}
    \vspace{-1em}
    \label{fig: meanLadders}
\end{figure}

To complement the numerical evaluation of the four different types of ladder, we provide in Fig.~\ref{fig: meanLadders} a visual representation of the average ladders based on the standard error over the whole dataset in the rate-quality-energy space. As anticipated, in the RQ domain the traditional rate-driven RQ-PF ladder exhibits the best performance (with tight standard error in both quality and bit rate dimensions) for the majority of ladder rungs. Nevertheless, this performance comes at the cost of higher energy expenditure. On the other hand, the two EQ-PF ladders exhibit better quality-rate-energy tradeoffs, particularly for the 50 to 90 VMAF range. Over that threshold all curves start converging or interlacing. The RE domain reveals the energy-to-bit cost~\cite{KatsenouPCS2022}, defined as the slope of the RQ line, which changes across different bitrate ranges. Furthermore, it is important to note that the quality-driven rule for the construction of the ladder appears to deliver ladders of lower bit rates and energy at equivalent high quality range. This opens the opportunity to apply the EQ-PF ladders to exploit both ``unnoticable'' as well ``acceptable''~\cite{Bingol_ICC2023}  differences in visual quality  in pursuit of sufficiency in the design of streaming services.

%%%%%%%%%%%%%%%%%%%%%%%%%%%%%%%%%%%%%%%%%%%%%%%%%%%%%%%%%%%%%%%%%
\section{Conclusions}
\label{sec:Conclusion}
This work investigated constructing bitrate ladders for adaptive streaming based on EQ curves rather than the conventional RQ curves. Encoding a subset of YouTube-UGC videos with x.265 showed substantial overlap in bitrates across resolutions but with shifted energy consumption, indicating potential for energy savings. Then, computing PFs from the EQ curves and using these to build ladders resulted in up to 31\% lower decoding energy for comparable quality levels at the cost of higher bitrate. The quality-driven ladder construction approach further reduced bitrates and energy. The results demonstrate the benefits of optimising for EQ over RQ in adaptive streaming ladder design. Future work will investigate these gains by adding the display device power consumption and in terms of carbon emission reductions. An improved expression of quality sufficiency will be investigated as well.

% \vfill\pagebreak
% \newpage
\bibliographystyle{IEEEbib}
\bibliography{references}

\end{document}